\documentclass[aps,prl,floatfix,twocolumn]{revtex4}
\usepackage{epsfig}
\begin{document}

%\twocolumn[
\hsize\textwidth\columnwidth\hsize\csname@twocolumnfalse\endcsname

\title{Magnetic bipolar transistor}
\author{Jaroslav Fabian}
\affiliation{Institute for Theoretical Physics, Karl-Franzens
University, Universit\"{a}tsplatz 5, 8010 Graz, Austria}

\author{Igor \v{Z}uti\'{c} and S. Das Sarma}
\affiliation{Condensed Matter Theory Center, Department of Physics,
University of Maryland, College Park, Maryland 20742}

\begin{abstract}

A magnetic bipolar transistor is a bipolar junction transistor with one or more
magnetic regions, and/or with an externally injected nonequilibrium (source) spin.  
It is shown that electrical spin injection through the transistor is
possible in the forward active regime. It is predicted that the current
amplification of the transistor can be tuned by spin.  
\end{abstract}
%\pacs{72.25.Dc,72.25.Mk}

\maketitle
\newpage

We propose~\cite{mbt2002} a novel device scheme---the magnetic bipolar transistor (MBT)---which builds
on the existing technology (bipolar junction transistor~\cite{Shockley:1950,Tiwari:1992}), adding spin degrees of
freedom to the current carriers. An MBT is a bipolar spintronic device: 
its functionality is defined by the transport properties of electrons, holes, and their
spins. While bipolar spintronics~\cite{Zutic2001:PRB,Zutic2001:APL,Zutic2002:PRL,
Fabian2002a:PRB} still relies rather on 
experimentally demonstrated fundamental physics concepts 
(such as spin injection~\cite{Fiederling1999:N,Ohno1999:N,Jonker2000:PRB,Hammar2001:APL}, 
spin filtering~\cite{Hao1990:PRB}, or 
semiconductor ferromagnetism~\cite{Ohno1992:PRL,Ohno1998:S,Saito2003:PRL})  
than on working devices, 
recent experiments~\cite{Kohda2001:JJAP,Johnston-Halperin2002:PRB} on spin injection
through bipolar tunnel junctions prove the potential of the spin-polarized bipolar 
transport for both fundamental physics and useful technological applications. 
Here we analyze MBTs 
(other types of spin transistors were proposed in 
Refs.~\onlinecite{Datta1990:APL,%
Monsma1995:PRL,Johnson1993:S,Ciuti2002:APL,vanDijken2003:PRL,Jiang2003:PRL,Bauer2003:APL}),
with a magnetic base and a source spin in the emitter.
We predict that spin can accumulate in the collector due to the electrical spin
injection, and that the current amplification 
of MBTs can be controlled by spin. 

Crucial to MBTs is the use of magnetic semiconductors where the 
splitting of the carrier bands produces spin-polarized electrons
or holes with the spin polarization perhaps 10\% or more. The
carrier band splitting can be of the Zeeman or the exchange
type. The former arises from large $g$ factors (for example,
in Cd$_{0.95}$Mn$_{0.05}$Se the $g$ factor exceeds 500~\cite{Dietl:1994b},
while it is as large as 50 in InSb at room temperature),
and an application of a magnetic field, while the latter comes 
from the exchange coupling in ferromagnetic semiconductors (about 10 meV).
In addition to the equilibrium spin, a nonequilibrium (source) spin can
be generated in the emitter by an external 
spin injection, electrical or optical\cite{Meier:1984}.

\begin{figure}
%\centerline{\psfig{file=new_schemes.eps,width=1\linewidth}}
\centerline{\psfig{file=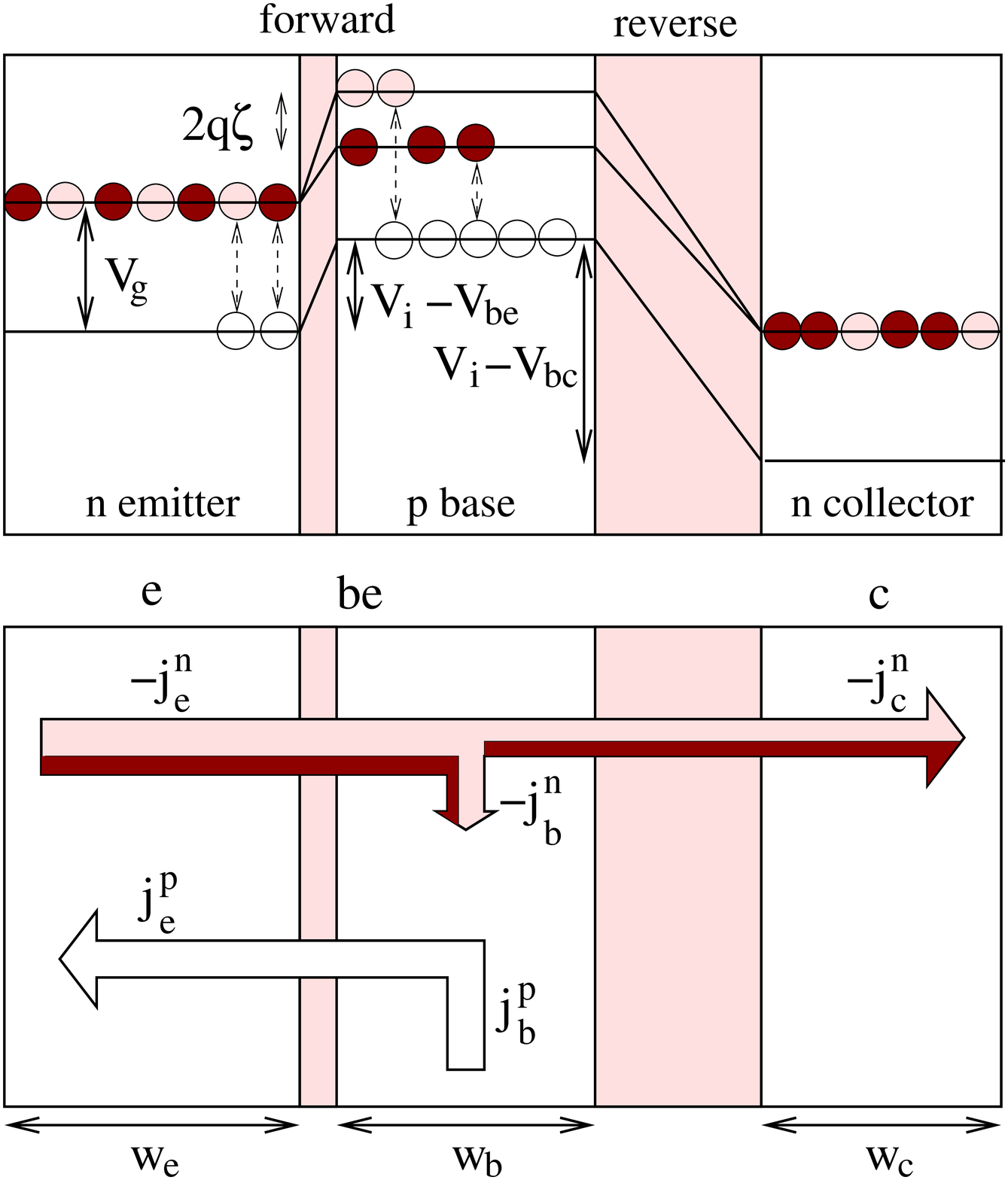,width=1\linewidth}}
\caption{Scheme of an $npn$ transistor with magnetic base. The top figure
shows the bands. The conduction band is separated by the
band gap $V_g$ from the valence band, and has a spin splitting (Zeeman 
or exchange) of 
$2 q \zeta$, 
leading to the equilibrium spin polarization
$\alpha_{0b}=\tanh(q \zeta/k_B T)$. Holes are unpolarized. 
The spin is indicated by the shade of the circles (dark
and light). The emitter-base junction is
forward biased with voltage $V_{be} > 0$ lowering the built-in voltage
$V_i$ and narrowing the depletion layer (shaded), while the base-collector
junction is reverse biased with voltage $V_{bc} < 0$, widening the 
depletion layer. Electrons flow easily from the emitter to the base,
where some of them recombine (dashed lines) with holes, the rest being
swept by the electric field in the base-collector depletion layer to the collector.
Holes, which are the large part of the base current, flow to the
emitter. The flow of electrons and holes is depicted ($j$ are
the corresponding charge currents) in the bottom
figure, where also the effective widths $w$ are indicated
and the regions labels ($e$, $be$, and $c$) for the densities are shown. 
The electron flow is spin-polarized, as indicated by the shading of the arrows.}
\label{fig:1}
\end{figure}

Our model is described in Fig.~\ref{fig:1}. We consider an $npn$ structure
doped with $N_{de}$ donors in the emitter, $N_{ab}$ acceptors in the base,
and $N_{dc}$ donors in the collectors. There are two depletion layers, one
between the emitter and the base, the other between the base and the collector.
The transistor is a three terminal device: there is a contact with an external
electrode at each region, generating bias $V_{be}$ across the emitter-base
and $V_{bc}$ across the base-collector depletion layer.  
The base is magnetic. For simplicity only electrons are spin polarized. 
The equilibrium spin polarization in the base is 
$\alpha_{0b}=\tanh(q \zeta/k_B T)$~\cite{Zutic2002:PRL},
where $2 q \zeta$ 
is the conduction band spin splitting and $k_B T$ is the thermal
energy. The nonequilibrium spin polarization injected externally into the
emitter is $\alpha_e$. The equilibrium number of electrons in the base
depends on the equilibrium spin polarization\cite{Fabian2002a:PRB}:
\begin{equation}
n_{0b}=(n_i^2/N_{ab})(1/\sqrt{1-\alpha_{0b}^2}),
\end{equation}
where $n_i$ is the intrinsic carrier density. The equilibrium
number of holes in the emitter is $p_{0e}=n_i^2/N_{de}$. For simplicity
we assume that the electron and hole diffusivities $D_n$ and $D_p$
are uniform, similarly for the electron and hole diffusion lengths $L_n$ 
and $L_p$, and for the electron spin diffusion length $L_s$. 
The effective widths $w$ (which depend on the biases as well as
on $\alpha_{0b}$~\cite{Fabian2002a:PRB}) of the three bulk regions  
are defined in Fig.~\ref{fig:1}. 

We consider the most useful forward active (also called amplification) regime of the transistor, 
where the emitter-base depletion
layer is forward biased, $V_{be} > 0$,
and the base-collector junction is reverse biased, $V_{bc}<0$, as shown in 
Fig.~\ref{fig:1}. 
Furthermore, we assume the small injection limit where the excess 
(injected) 
electron densities anywhere in the structure are smaller than the 
equilibrium densities determined by the doping.  
The resulting flow of electrons
and holes is depicted in the bottom part of Fig.~\ref{fig:1}. Take electrons
first. As the barrier between the emitter and the collector is lowered
by $V_{be}$, the electrons flow easily to the base, forming the electron
emitter current $j_e^n$. In the base the excess electrons
either recombine with holes, producing the base recombination current  
$j_b^n$, or diffuse towards the base-collector depletion layer. This layer
is reverse biased so that all the electrons reaching it from the base are
swept by the large electric field 
to the collector, forming the collector current $j_c^n$.
Holes need to be supplied from the base to go in the forward direction
to the emitter, forming the hole base,  $j_b^p$, and
the hole emitter, $j_e^p$, currents. The total emitter current is
$j_e=j_e^n+j_e^p$ and the total collector current is $j_c=j_c^n$. 
The base current is $j_b=j_e-j_c$. The current amplification coefficient
(gain) 
is defined as $\beta=j_c/j_b$, being about 100 for practical 
transistors: for a small variation in $j_b$ (input 
signal), there is a large variation in $j_c$ (output signal). In
the following we show that the electron flow in MBTs brings about
spin accumulation (nonequilibrium spin) in the collector, proving
the possibility of the electrical spin injection. We also 
show that $\beta$ depends on both $\alpha_e$ 
and on $\alpha_{0b}$, predicting a spin control of the gain.

\paragraph{Electrical spin injection.} Our goal is to 
calculate how much spin polarization $\alpha_c=s_c/N_{dc}$ 
will accumulate in the collector in response to the nonequilibrium
spin $\alpha_e$ and the equilibrium spin 
$\alpha_{0b}$. 
To that end we model the emitter-base junction as a forward
biased magnetic {\it p-n} junction with equilibrium spin polarization $\alpha_{0b}$
in the $p$ region (base)  and source spin polarization $\alpha_c$ in
the $n$ region (emitter). Our theory of magnetic {\it p-n} junctions ~\cite{Fabian2002a:PRB}
determines the electron $n_{be}$ and spin $s_{be}$ densities at the $be$ region (see 
Fig.~\ref{fig:1})
at the depletion layer: 
\begin{eqnarray} \label{eq:nbe}
n_{be} = & n_{0b} e^{qV_{be}/k_B T} \left (1+\alpha_e \alpha_{0b} \right ), \\
\label{eq:sbe}
s_{be} = & n_{0b} e^{qV_{be}/k_B T} \left (\alpha_{0b}+ \alpha_e  \right ).
\end{eqnarray}
The nonequilibrium electron density injected into the base depends on the
product $\alpha_e\alpha_{0b}$, realizing the Silsbee-Johnson
spin-charge coupling~\cite{Silsbee1980:BMR,Johnson1985:PRL}.
If $\alpha_e=0$, Eq.~(\ref{eq:nbe}) reduces to the standard 
Shockley's equation~\cite{Shockley:1950} for the nonequilibrium
minority electron density in a biased {\it p-n} junction.

We next model the base-collector junction as another magnetic {\it p-n}
junction. This junction is reverse biased, and has {\it both} the
equilibrium spin polarization $\alpha_{0b}$ and the source spin density 
$s_{be}$ in the $p$ region (base). This is the case of a magnetic solar
cell~\cite{Zutic2001:APL}, since the electron and the source spin densities in the $p$ region 
mimic the carrier and spin generation by light. For this case our
theory \cite{Fabian2002a:PRB} gives
\begin{equation} \label{eq:sc}
s_{c}\approx \gamma_1 s_{be} = \gamma_1 n_{0b} e^{qV_{be}/k_BT}
\left (\alpha_{0b} +\alpha_e \right ),
\end{equation}
where $\gamma_1\approx \left (L_s/w_b \right )\tanh\left (w_c/L_{s}\right)$.
The accumulated spin polarization, which is the measure of the electrical
spin injection efficiency, is $\alpha_c=s_c/N_{dc}$. 
Typically the spin diffusion length in the collector $L_{sc} \gg w_b$, which means that
 $\alpha_c$ can be a considerable fraction (say, 10\%) of $\alpha_e$ or $\alpha_{0b}$. What is
interesting in Eq.~(\ref{eq:sc}) is the fact that $\alpha_{0b}$ plays
the same role as $\alpha_e$ in the spin injection: the equilibrium 
spin can cause spin accumulation in the low injection limit, because
it leads first to nonequilibrium spin $s_{be}$.  
This has no analog in magnetic diodes, where spin accumulation cannot
result from the presence of just an equilibrium spin polarization.

\paragraph{Spin control of current amplification.} 
When written in
terms of $n_{be}$, the formulas for the currents $j_e$ and $j_c$
are the same as for the standard (nonmagnetic) bipolar transistors
derived by Shockley~\cite{Shockley:1950}. After we write those formulas for the active
forward regime, we substitute Eq.~(\ref{eq:nbe}) for $n_{be}$
and obtain the dependence of the currents (and of $\beta$) on $\alpha_e$ and
$\alpha_{0b}$.
 
The emitter current is
\begin{equation} \label{eq:je}
j_e=j_{gb}^n (n_{be}/n_{0b})+j_{ge}^p (p_{e}/p_{0e}),
\end{equation}
where $j_{gb}^n$ is the electron generation current 
$(qD_n/L_n) n_{0b}\coth(w_b/L_n)$, the hole generation 
current $j_{ge}^p$ is $(qD_p/L_p) p_{0e}\coth(w_e/L_p)$,
and the injected hole density in the emitter
$p_e=p_{0e}\exp(qV_{be}/k_BT)$. 
The collector current comprises only electrons (Fig.~\ref{fig:1}):
\begin{equation} \label{eq:jc}
j_c=j_{gb}^n (n_{be}/n_{0b})\cosh(w_b/L_n). 
\end{equation}
After evaluating 
$j_b=j_e-j_c$ and substituting Eq.~(\ref{eq:nbe}) for $n_{be}$, 
it is straightforward to 
show that in the narrow base limit ($w_b \ll L_n, L_s$) the gain
is 
\begin{equation}
\beta=1/(\alpha_T'+\gamma'),
\end{equation}
where we use the standard transistor notation~\cite{Tiwari:1992}: 
\begin{eqnarray}
\alpha_T'&=&(w_b/L_n)^2/2, \\
\gamma'&=&\frac{N_bD_p}{N_eD_n}\frac{w_b}{L_{p}\tanh(w_e/L_{p})}
\frac{\sqrt{1-\alpha_{0b}^2}}{\left (1+\alpha_e \alpha_{0b}\right )}.
\end{eqnarray}
The factor $\alpha_T'$ determines how much electrons will recombine in the base, 
thus not reaching the collector. This factor is not affected by the 
presence of spin, and is the same as in the standard transistors.
The factor $\gamma'$ is related to the emitter injection efficiency,
since it measures the proportion of the electron flow in the emitter
current (where both electrons and holes contribute). This factor
does depend on the spin.
To get the maximum amplification, both $\alpha_T'$ and $\gamma'$ need to 
be small. For the most efficient spin control of $\beta$, one
needs $\alpha_T' \alt \gamma'$, the case of Si-based
transistors which have slow carrier recombination. In this case 
\begin{equation} \label{eq:gain}
\beta(\alpha_e,\alpha_{0b})=\beta(\alpha_e=0,\alpha_{0b}=0) \times
\frac{1+\alpha_e\alpha_{0b}}{\sqrt{1-\alpha_{0b}^2}} .
\end{equation}
The current amplification is affected by both $\alpha_e$ and $\alpha_{0b}$.

\begin{figure}
\centerline{\psfig{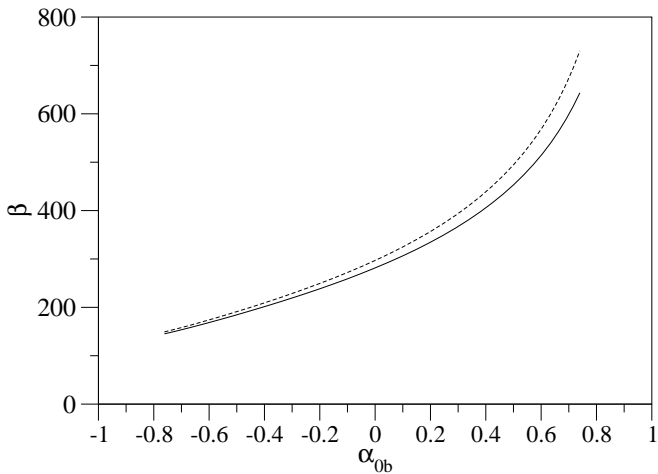}}
\caption{Calculated current amplification coefficient $\beta$  as a function of $\alpha_{0b}$ for
a fixed $\alpha_e$. The dashed line is Eq.~(\ref{eq:gain}).
}
\label{fig:2}
\end{figure}

As an illustration we 
calculate $\beta$ as a function of $\alpha_{0b}$ for an MBT 
with $\alpha_e=0.9$ and with generic materials parameters, specified
for a Si-like transistor.
The nominal widths of the emitter, base, and collector
are 2, 1.5, and 2 $\mu$m, respectively. The dopings are $N_e=10^{17}$, $N_b=10^{16}$,
and $N_c=10^{15}$ cm$^{-3}$. Electron (hole) diffusivities at room temperature
are taken to be $D_n=100$ and $D_p=10$ cm$^2$/s. The bias
voltages are $V_{be}=0.5$ and $V_{bc}=0$ volts. 
The intrinsic carrier density $n_i=10^{10}$ cm$^{-3}$ and the 
dielectric constant (needed to calculate the effective widths $w$) is 12. 
The carrier and spin diffusion lengths (note that Si has long
recombination and spin relaxation times~\cite{Lepin1970:PRB}) are taken to be
$L_n=30$ $\mu$m, $L_p = L_s = 10$ $\mu$m.  The calculated $\beta$ varies
strongly with the spin, following closely the approximate $\beta$ given
by Eq.~(\ref{eq:gain}). The amplification is largest (smallest) for the parallel (antiparallel)
orientation of the source and equilibrium spins. 

We conclude that spin can be injected through MBTs and that current
amplification can be controlled by both the source and the equilibrium spin, 
making MBTs attractive for spintronic applications.

This work was supported by DARPA, the NSF-ECS, and the US ONR. 

\bibliographystyle{apsrev}
\bibliography{spintronics}

\end{document}